# New Field Model of Polymer/Nanoparticle Mixture

# —Realizing Discreteness in the Continuous Description

Hui-shu Li[1], Bo-kai Zhang[2,1], Jian Li[3,1], Wen-de Tian[1,4,*] and Kang Chen[1,4,*]

[1]Center for Soft Condensed Matter Physics & Interdisciplinary Research, College of Physics, Optoelectronics and Energy, Soochow University, Suzhou 215006, China.

[2]National Laboratory of Solid State Microstructures and Department of Physics, Nanjing University, Nanjing 210093, China

[3]Department of Physics, Nanjing Normal University, Nanjing 210023, China

[4]Kavli Institute for Theoretical Physics China, CAS, Beijing 100190, China

[*]Authors to whom Correspondence should be addressed. Electronic mail: kangchen@suda.edu.cn (K.C.); tianwende@suda.edu.cn (W.d.T.)

## Abstract

Field-theoretical method is efficient in predicting the assembling structures of polymeric systems. However, for the polymer/nanoparticle mixture, the continuous density description is not suitable to capture the realistic assembly of particles, especially when the size of particle is much larger than the polymer segment. Here, we developed a *field-based* model, in which the particles are eventually discrete and hence it can overcome the drawbacks of the conventional field descriptions, e.g., inadequate and crude treatment on the polymer-particle interface and the excluded-volume interaction. We applied the model to study the simplest system of nanoparticles immersed in dense homopolymer solution. Our model can address the depletion effect and interfacial interaction in a more delicate way. Insights into the enthalpic and/or entropic origin of the structural variation due to the competition between depletion and interfacial interaction are obtained. New phenomena such as depletion-enhanced bridging aggregation are observed in the case of strong interfacial attraction and large depletion length. This approach is readily extendable to studying more complex polymer-based nanocomposites or biology-related systems, such as dendrimer/drug encapsulation and membrane/particle assembly.

## I. INTRODUCTION

Addition of nanofillers to polymer materials has long been a practical approach to fabricate flexible composites with enhanced mechanical, electrical or optical properties.[1-6] Understanding and controlling the formation of the assembling structures of polymer nanocomposites (PNCs) are the keys to designing and realizing the desired macroscopic performance. Many theoretical and simulation methods have been developed[7-31] to study the assembling behaviors of PNCs. The self-consistent field theory (SCFT) is one of the powerful approaches to investigate the mesoscopic structures of multicomponent polymeric systems.[32-34] Hence, efforts have been made to establish a field-based model for PNCs.[18, 22, 29, 30] The challenges facing the incorporation of particles into the field-based model of polymers are how to treat the interface between the polymers and the particles and how to address the strong excluded-volume (EV) interactions both between the polymers and the particles and between the particles. The hybrid theory proposed by Thompson and coworkers couples the SCFT for the polymers with the density functional theory for the particles. This hybrid method has obtained many successes in predicting the ordered structures of diblock copolymer/nanoparticle composites.[18, 30, 35-37] But it's well understood that the EV and the interfacial interactions between the polymers and the particles are not appropriately considered in this approach. Another type of hybrid method combines the field description of the polymers and particle-based Brownian dynamics (BD) simulation of the particles.[22, 38, 39] The explicit particle coordinates potentially enable the method to overcome the challenges mentioned above. However, the coupling between the SCFT iterations and BD motions becomes a tricky problem. One major difficulty is how to determine the force on the particles exerted by surrounding polymers. Sides and coworkers[22] use the *explicit partial derivative* of the Hamiltonian to the particle positions as the coupling force on the particles, which is questionable in the sense of the quasi-static approximation, i.e. the polymer matrix is in "equilibrium" between two successive steps of BD motion of the particles. Note that the auxiliary fields ($\omega$'s) in SCFT are implicit functions of the particle positions. Total derivative of the Hamiltonian to the particle positions is more appropriate instead, while the expression of which is not available. This issue may not affect the result seriously since the most concerned aspect is the equilibrium assembling structure, not the dynamics. Another issue of the SCFT/BD method is that the alternating SCFT iterations and small-step BD motions make it very computational. To avoid the complication of considering the EV interactions in the polymer/multi-nanoparticle assembling system, some works focused on the dilute limit (one or two fixed particles)[28, 40] or made a simplification by invoking the concept of effective polymer concentration.[41]

Here, we developed a field-based model of PNCs. Unlike the conventional field models[18, 29, 30] in which particles are described by continuous density distribution, the particles in our model are eventually discrete as in the particle-based description, i.e. the particle density in our model no longer represents the ensemble-averaged density distribution of particles. The goal of our model is to obtain the ensemble-averaged density distributions of the polymers with discrete nanoparticles located at the "*most probable*" positions, hence the structure we predict is more like an instantaneous pattern of the system. The motivations behind are three folds: 1) patterns with instantaneous locations of particles can better exhibit the delicate structures (especially when the particles are packed); 2) discrete description of particles allows delicate treatment on the EV interactions and the polymer-particle interface; 3) due to 2), the evaluation on the polymer-particle interfacial energy is more appropriate and the contribution of the depletion effect can also be involved. We applied this model to study the simplest system of nanoparticles immersed in dense homopolymer solution or melt. We focused on the structural variations under diverse polymer-particle interfacial interactions and depletion lengths.

The paper is organized as follows: we describe the basics of our model and method in Sec. II and more detailed equations are given in the Appendix. Results are analyzed and discussed in Sec. III. We briefly conclude in Sec IV.

## II. MODEL AND METHOD

We consider a system consisting of a mixture of $n_p$ (variable) homopolymers and $n_c$ spherical solid particles. The radius of all particles is $R_c$. The coarse-grained polymer chain is composed of N segments of size $\sigma$. The bonded interactions of Gaussian chain are quantified by an elastic potential energy:

$$\beta U_0\left[\vec{R}(s)\right] = \frac{1}{4R_{g0}^2}\sum_{i=1}^{n_p}\int_0^1 ds\left|\frac{d\vec{R}_i(s)}{ds}\right|^2 \quad (1)$$

where $\beta = 1/k_BT$, $\vec{R}(s)$ represents the configuration of the polymer chain, $R_{g0}$ is the unperturbed radius of gyration. We employ the quadratic compressible model to address the EV interaction between segments with the presence of particles:

$$\beta U_p = \sigma^{-d}\int d\vec{r}\,\frac{\kappa^{-1}}{2}\left[\phi_p(\vec{r}) + \phi_c^{eff}(\vec{r}) - 1\right]^2\left\{H\left[\phi_{hc} - \phi_c(\vec{r})\right] + H\left[\phi_c(\vec{r}) - \phi_{hc}\right]H\left[1 - \phi_p(\vec{r}) - \phi_c(\vec{r})\right]\right\}$$

(2)

$d$ is the dimension, $\kappa$ is a dimensionless parameter proportional to the compressibility of the polymer matrix. $\phi_p(\vec{r})$ and $\phi_c(\vec{r})$ are dimensionless concentrations of segments and particles, respectively. $\phi_c(\vec{r}) = \int d\vec{r}'\, 0.5\rho_c(\vec{r}+\vec{r}')\left\{1-\tanh\left[3\left(\left|\vec{r}'\right|-R_c\right)/\sigma\right]\right\}$, where $\rho_c(\vec{r})$ is the number density of particles whose surface profile is smoothed by the hyperbolic tangent function for numerical efficiency. *Note that, principally, the concentration and density quantities in the above and the following energy expressions, etc. should be operators or instantaneous ones. Instead, we directly use the "ensemble-averaged" ones for simplicity since they are interchangeable with instantaneous ones in the mean field approximation.*

$\phi_c^{eff}(\vec{r})$ is introduced to take into account the depletion layer surrounding the surface of each particle and its capability of overlapping with each other and with the entity of other particles, which causes the so-called depletion effect (see Fig.1). We approximate that the density profile of polymers surrounding a particle is described by a hyperbolic tangent function in case of neutral polymer-particle interfacial interaction. Then, the particle at position $\vec{r}'$ generates an effective particle concentration which is felt by polymers at position $\vec{r}$:

$$\phi_{c,\vec{r}'}(\vec{r}) = 0.5l^d\rho_c(\vec{r}')\left\{1-\tanh\left[3\left(\left|\vec{r}-\vec{r}'\right|-R_c-\xi_D\right)\big/\left(\xi_D+\sigma\right)\right]\right\}H\left(\left|\vec{r}-\vec{r}'\right|-R_c\right) \quad (3)$$

where $l$ is the grid size of the simulation box. The width of the depletion layer, $\xi_D$, is related to the rigidity of the polymer chain. The Heaviside function in Eq. (3) guarantees the position $\vec{r}$ is outside the entity of the particle at position $\vec{r}'$. The effective particle concentration is then:

$$\phi_c^{eff}(\vec{r}) = max\left\{\phi_c(\vec{r}), \max_{\vec{r}'}\left[\phi_{c,\vec{r}'}(\vec{r})\right]\right\} \quad (4)$$

The first max function addresses the overlap between the depletion layer and the entity of particles, while the second one addresses the overlap between depletion layers of nearby particles. Figure 1 shows the concentration plots of two contact dicrete particles in our model. Apparently, the particles are surrounded by a depletion layer mimicked by effective particle concentration (or spatial occupation) felt by polymers. And, the depletion layers can overlap with each other and with the entity of particles.

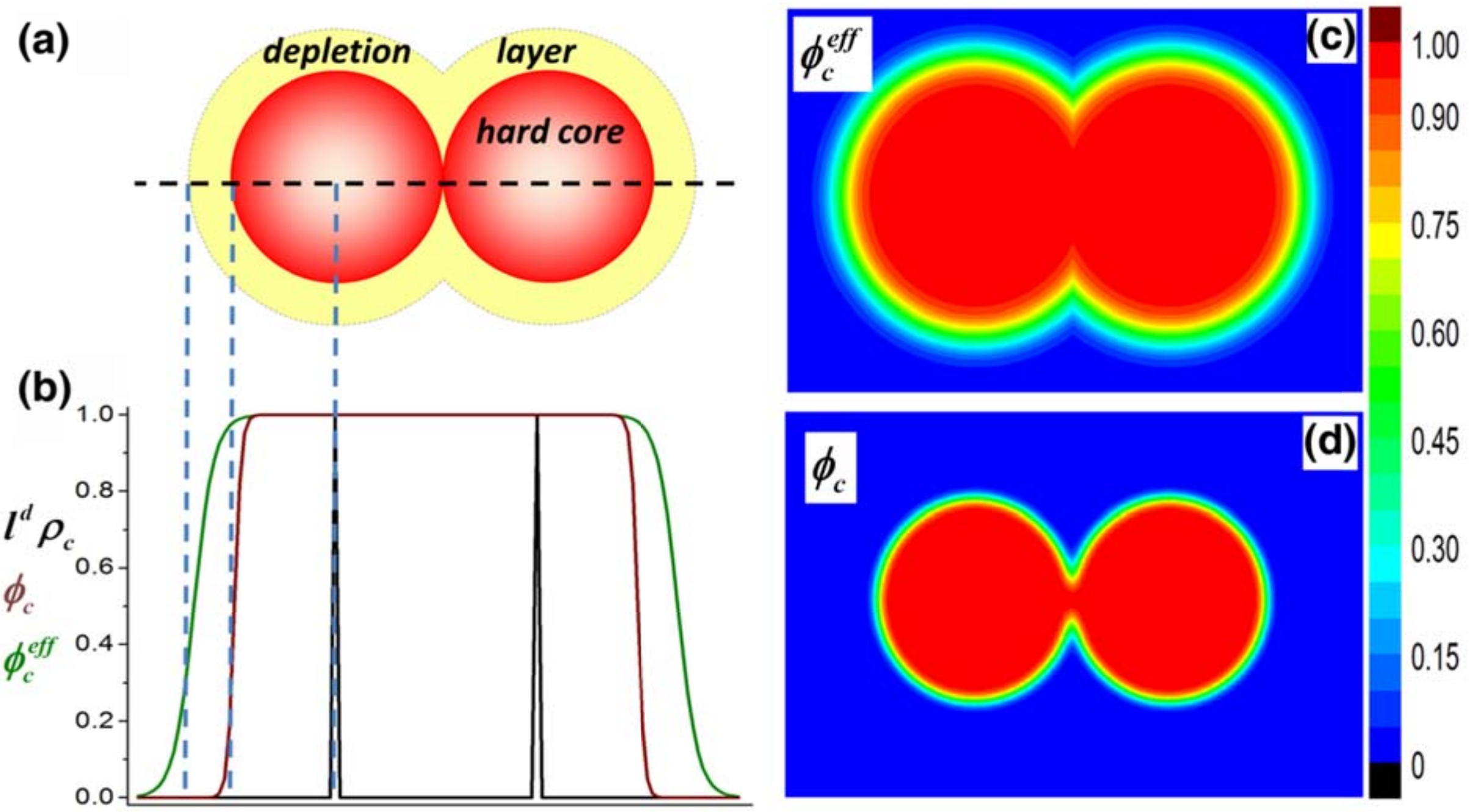

**FIG 1.** Schematic and concentration plots of two complete particles in contact in two dimensions. ($\xi_D = 0.4R_c$.) (a) Schematic of the two particles shows the hardcore regions and depletion layers. (b) Concentration profiles of $l^d \rho_c(\vec{r})$ (black), $\phi_c(\vec{r})$ (red) and $\phi_c^{eff}(\vec{r})$ (green) along the center-to-center line of two particles. (c) Concentration plot of $\phi_c^{eff}(\vec{r})$. (d) Concentration plot of $\phi_c(\vec{r})$.

To address the strong polymer-particle EV interaction and differentiate it from the polymer-polymer EV interaction, $\phi_{hc}$ in Eq. (2) is introduced as a threshold quantity to decide whether a location of the system is in the hardcore or non-hardcore regions of particles. We set this threshold quantity $\phi_{hc} = 0.3 \max_{\bar{r}}\left[\phi_c(\bar{r})\right]$, i.e., about one third of the maximum particle concentration in the system. $\phi_{hc}$ varies as the particles gradually become discrete during the calculation. Eventually, $\phi_{hc} = 0.3$. The choice of 0.3 is empirical; other choices of this number only quantitatively influence the results. The Heaviside functions in the curly braces of Eq. (2) pick the non-hardcore regions or the hardcore regions where the total concentration is less than one. In these two cases, only the weak polymer-polymer EV interaction characterized by $\kappa^{-1}$ is considered

The strong polymer-particle EV interaction is triggered where the total concentration in the hardcore region of particles is larger than one (the first term of Eq.(5)):

$$\begin{aligned}\beta U_{EV} &= \sigma^{-d}\int d\bar{r}\,\frac{\kappa_h^{-1}}{2}\left[\phi_p(\bar{r})+\phi_c(\bar{r})-1\right]^2 H\left[\phi_p(\bar{r})+\phi_c(\bar{r})-1\right]H\left[\phi_c(\bar{r})-\phi_{hc}\right]\\ &+\sigma^{-d}\int d\bar{r}\int d\bar{r}'\rho_c(\bar{r})\frac{\kappa_h^{-1}}{2}V_{op}\left(\left|\bar{r}-\bar{r}'\right|\right)\rho_c(\bar{r}')\end{aligned} \quad (5)$$

The second term of Eq. (5) represents particle-particle hardcore repulsion which is set to be proportional to the overlap "volume", $V_{op}$, between two nearby particles. $\kappa_h^{-1}$ is chosen to be large enough to avoid overlap between particles and between polymer and particle. *Note that even though there is no problem to express the strong EV repulsions as in Eq. (5) in terms of instantaneous concentrations, it is not appropriate to replace them by the "ensemble-averaged" ones in the conventional mean field approach. As mentioned in the introduction, the particle density in our model is of instantaneous nature instead of ensemble averaged which allows us to use Eq. (5) to address the strong EV interactions.*

The chemical nature of the polymers and the particles is encoded in an interfacial energy of exponential form:[25]

$$\beta U_s = \sigma^{-d} \int d\vec{r} \int_{R_c \le |\vec{r}' - \vec{r}|} d\vec{r}' \; \varepsilon \exp\left[ -\left( |\vec{r}' - \vec{r}| - R_c \right) / \Delta \right] \rho_c(\vec{r}) \phi_p(\vec{r}') \tag{6}$$

$\varepsilon$ is the dimensionless strength of the interfacial interaction or relative affinity between polymer and particle (positive for repulsion and negative for attraction). $\Delta$ denotes the spatial range. Besides the above potentials of real origins, an artificial double-well like potential is introduced to force the formation of discrete particles:

$$\beta U_A = \sigma^d \int d\vec{r} \frac{\lambda}{2} \left\{ \rho_c(\vec{r})^2 H\left[ \bar{\rho}_c - \rho_c(\vec{r}) \right] + \left[ \rho_c(\vec{r}) - l^{-d} \right]^2 H\left[ \rho_c(\vec{r}) - \bar{\rho}_c \right] \right\} \tag{7}$$

$\lambda$ is the strength of the two harmonic potentials which drive the particle number density at each grid in the simulation box toward 0 or $l^{-d}$. This artificial potential embodies the inseparability of a real particle. Note that this potential is position-unbiased, i.e., it does not directly influence the spatial arrangements of particles.

We fix $R_{g0} = 4.08\sigma$. The chemical potential of polymers, $\mu$, is chosen that the concentration of bulk polymers is 1. We set $\kappa^{-1} = 3.33$ which corresponds to the compressibility of polymethylmethacrylate melt at 450K.[42, 43] The SCFT calculations were performed in two dimensions (see Appendix).[44, 45]

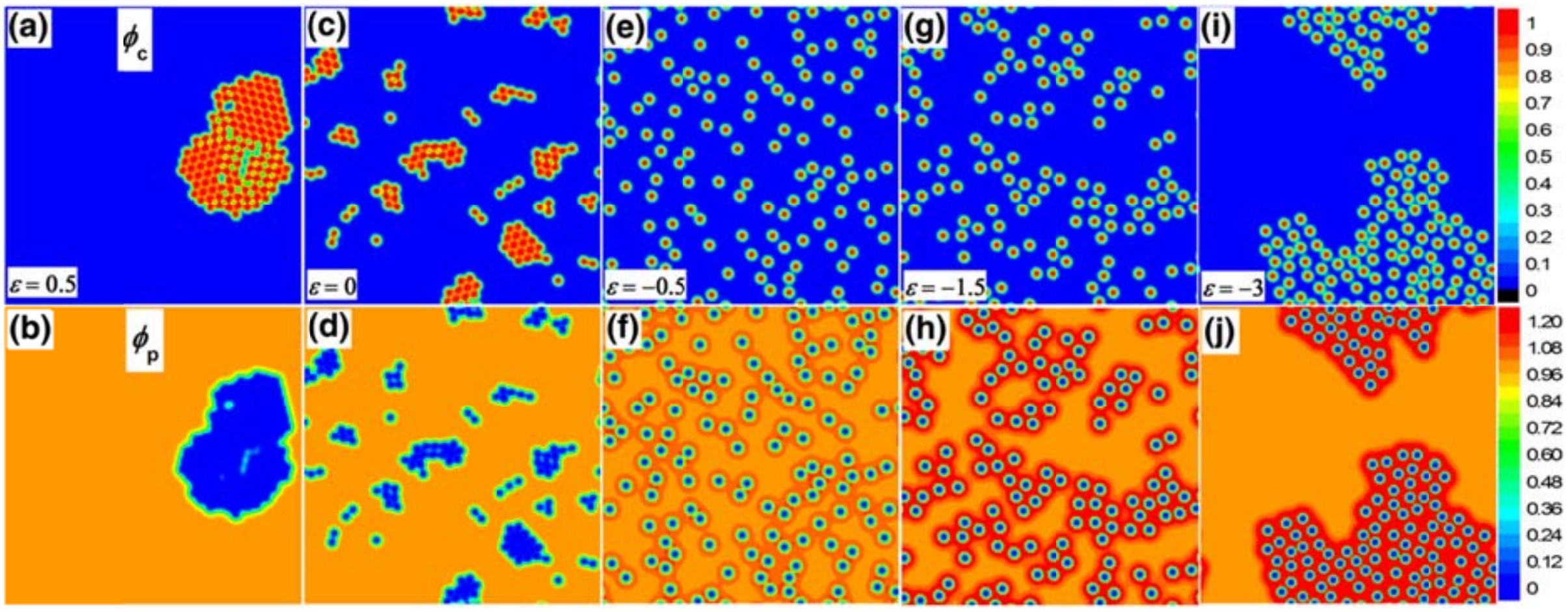

**FIG 2.** Concentration plots of nanoparticles, $\phi_c(\vec{r})$ (top row) and polymers, $\phi_p(\vec{r})$ (bottom row) in two dimensions. There are 130 small nanoparticles ($R_c = \sigma \approx 0.24 R_{g,0}$) immersed in a polymer matrix with dimensions of $15.7 R_{g0} \times 15.7 R_{g0}$ (average area fraction of particles, $\bar{\phi}_c \approx 0.1$). $\Delta = 0.5\sigma$ is fixed. The strength varies: $\varepsilon$ = 0.5 (a and b); 0 (c and d); -0.5 (e and f); -1.5 (g and h) and -3 (i and j).

## III. RESULTS AND DISCUSSION

### A. Assembly of small nanoparticles

We first considered the case of small nanoparticles ($R_c = \sigma \approx 0.24 R_{g,0}$). The depletion effect is neglected (by setting $\xi_D = 0$), since the size of the nanoparticles is comparable to the segment. The assembly structures of varying the strength of interfacial interaction are shown in Fig. 2, (a)-(j). The particles in these final morphologies are discrete as in the particle-based simulations, showing clear interfaces with surrounding polymers. The packing or dispersion of these particles is explicit. For weak repulsion, $\varepsilon = 0.5$ (Fig. 2(a) and (b)), the particles macroscopically separate from the polymer matrix and aggregate contactly (closely packed) into one large cluster (Contact Aggregation, CA) to minimize the interfacial energy. It is a typical enthalpic-driven phase separation of a binary mixture. Nonuniform clusters of contact-aggregated particles (Contact-Aggregation Clusters, CAC) are formed for very small repulsion (or $\varepsilon = 0$, where weak entropic attraction is caused by the hyperbolic tangent surface profile of particles) (Fig. 2(c) and (d)). As expected,[26] in the case of weak attraction (Fig. 2(e) and (f)), particles are dispersed randomly in the polymer matrix, each one of which is coated with a layer of slightly more concentrated polymers (Random Dispersion, RD). As attraction is enhanced, the bound polymer layers get denser and particles get closer to share the bound layers. This apparent aggregation of particles is ascribed to the bridging effect.[26] Irregular domains rich in both polymers and particles (Loosely Bridging Aggregation, LBA) are formed (Fig. 2(g) and (h)) at intermediate attraction strength. For strong attraction ($\varepsilon = -3.0$), the nanoparticles aggregate closely (Closely Bridging Aggregation, CBA). The system can be viewed as composed of two separated phases: bulk concentrated polymer solution or melt and the concentrated phase rich in both polymers and particles (Fig. 2(i) and (j)).

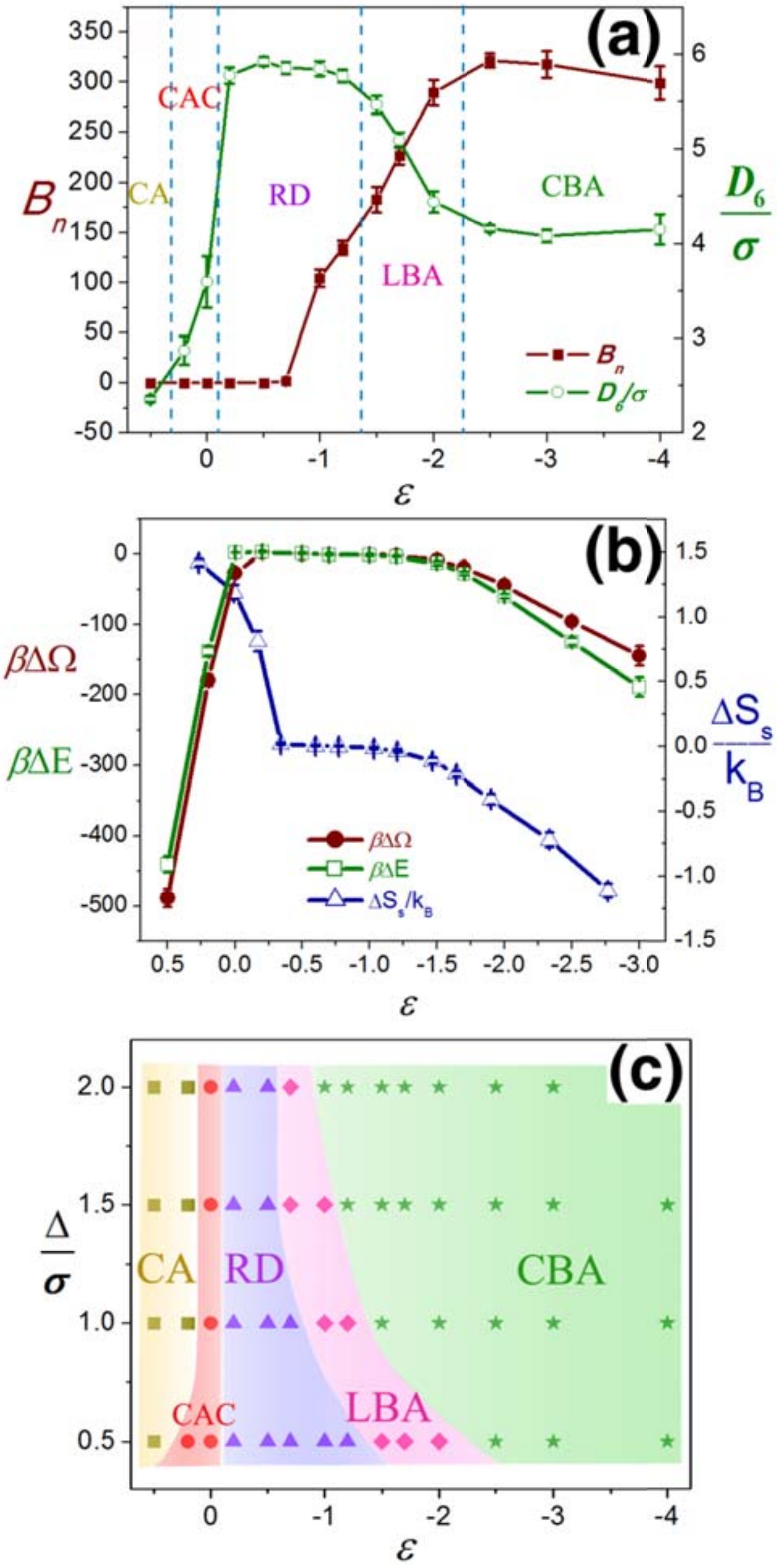

**FIG 3.** Analysis of the structural variation of small nanoparticles immersed in polymer matrix. (The system parameters are the same as in Fig. 2. Every data point is averaged over 10 independent runs.) (a) $B_n$ (red solid squares) and $D_6$ (green open circles) as functions of $\varepsilon$ ($\Delta = 0.5\sigma$). (b) The particle distribution of Fig. 2e is taken as the reference. The curves show the differences of grand potential ($\Delta\Omega$, red solid circles), potential energy ($\Delta E$, green open squares) and entropy per polymer chain ($\Delta S_s$, blue open triangles) between the "equilibrium" and the reference particle distributions ($\Delta = 0.5\sigma$). (c) Phase diagram in the $\Delta - \varepsilon$ space.

## B. Phase diagram and thermodynamic analysis.

To quantitatively analyze the structures of polymer/nanoparticle composites in Fig. 2, we introduce two useful quantities: the number of pairs of bridging-connected particles, $B_n$ and the mean distance between a particle and its six nearest neighbors, $D_6$. $B_n$ reflects the degree of bridging aggregation of particles from the view of amount. $D_6$ reflects the average degree of packing of particles from the view of distance. Two particles are deemed bridging-connected if a) their surface-to-surface distance is less than $4\sigma$, and b) the concentration of polymers in between exceeds 1.2. The choice of $4\sigma$ is based on the calculation of two-particle potential of mean force in the dilute particle limit, which shows the range of attraction well for $\varepsilon < 0$ extends to $4\sigma$. The number 1.2 (>1) is empirical, which does not influence the trend of $B_n$. The variations of these two quantities with $\varepsilon$ are shown in Fig. 3(a). Every data point is averaged over 10 independent runs. When $\varepsilon$ decreases from 0.5 to 0, $D_6$ grows rapidly due to the increase of the number of particles at the interface (during the variation of the structure from CA to CAC). A sharp transition happens between $\varepsilon = 0$ and $\varepsilon = -0.2$ that particles are no longer in contact (i.e., the transition of the structure from CAC to RD). For weak attraction ($-1 < \varepsilon < 0$), $B_n \simeq 0$ and $D_6$ reaches a maximum plateau, which reveals the well dispersion of particles in the polymer matrix. As the enhancement of the attraction, $B_n$ ($D_6$) increases (decreases) gradually, indicating the formation of more and more bridging connections or the closer and closer aggregation of particles (i.e., the formation of LBA). At very large attraction, $B_n$ ($D_6$) saturates at ~300 pairs of bridging connections (average center-to-center distance of $\sim 4\sigma$) which corresponds to the CBA structure. The calculations of $B_n$ and $D_6$ help determine the phase diagram in Fig. 3©.

To thermodynamically understand the variation of the structures, we took the particle distribution in RD phases, specifically the distribution of particles in Fig. 2(e), as the reference. We calculated the differences of grand potential ($\Delta\Omega$), potential energy ($\Delta E$) and entropy per polymer chain ($\Delta S_s$) between the "equilibrium" and the reference structure at various $\varepsilon$ (Fig. 3(b)). For $\varepsilon > 0$, $\Delta E$ is negative and $\Delta S_s$ is positive, i.e., both the enthalpy and the entropy are the driving force for the formation of contact aggregation of particles. The values of $\Delta E$ and $\Delta\Omega$ are close, which means enthalpy is the main driving force. For $\varepsilon = 0$, the potential energy does not change ($\Delta E = 0$) with the rearrangement of particles, the negativity of $\Delta\Omega$ comes from the increase of the entropy of polymer chains, i.e., the CAC structure at $\varepsilon = 0$ is solely entropic-driven. When $\varepsilon < -1$, both $\Delta E$ and $\Delta S_s$ are negative, i.e., enthalpy (entropy) favors (disfavors) the bridging aggregation of particles. Figure 3(c) shows the phase diagram in the $\varepsilon - \Delta$ space. The regions of the intermediate structures, CAC and LBA, are narrow. With the increase of $\Delta$, the regions of CAC, LBA and RD all shrink, manifesting that long-range interfacial interaction suppresses the dispersion of small nanoparticles in the polymer matrix.

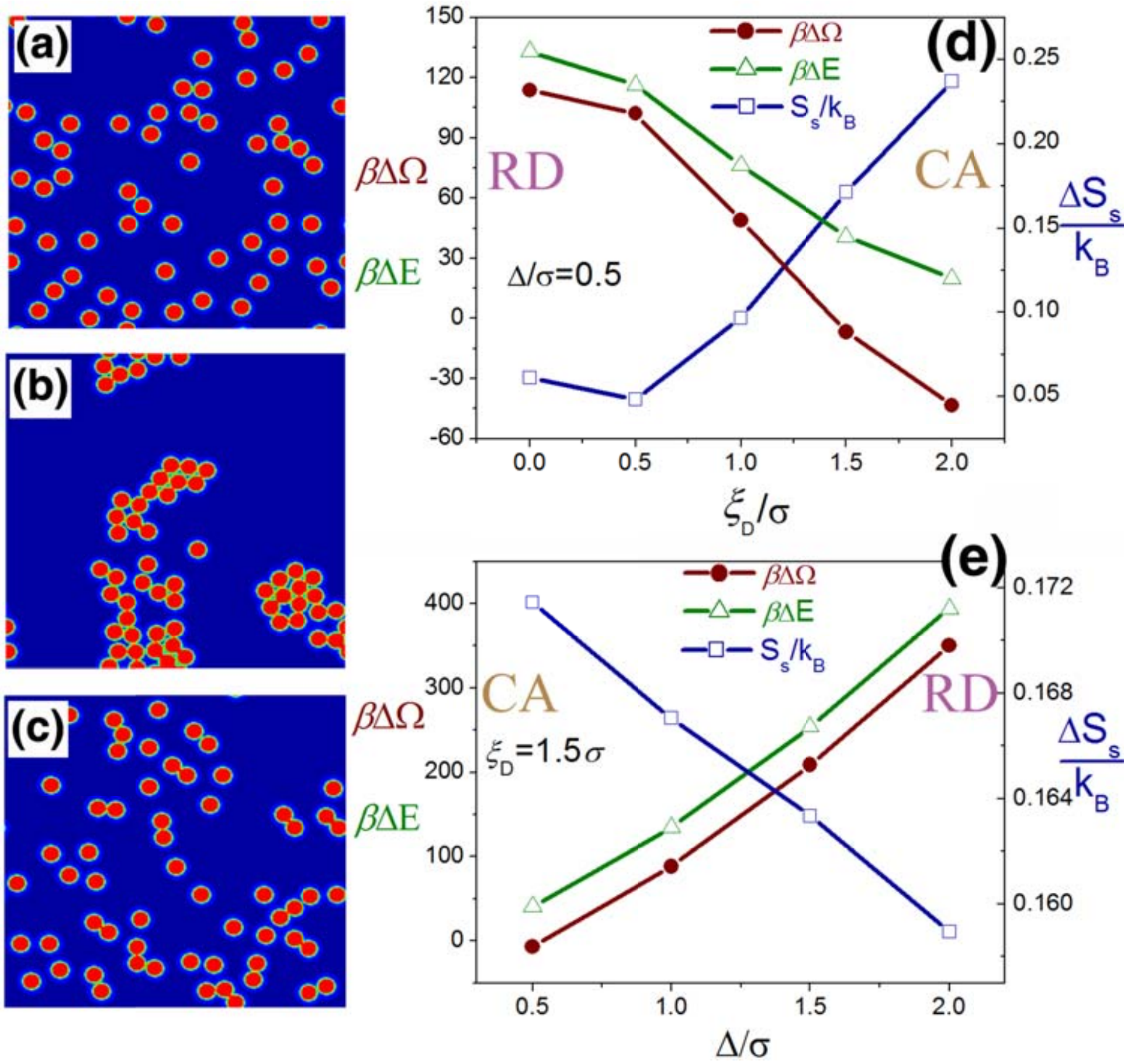

**FIG 4.** Structural variation of large nanoparticle/polymer composites in the case of weak attraction ($\varepsilon = -0.2$). There are 60 nanoparticles ($R_c = 5\sigma \approx 1.22 R_{g,0}$) immersed in the polymer matrix with the dimensions of $47R_{g0} \times 47R_{g0}$ ($\bar{\phi}_c \approx 0.13$). Three concentration plots of nanoparticles are for (a) $\Delta = 0.5\sigma$, $\xi_D = 0$; (b) $\Delta = 0.5\sigma$, $\xi_D = 1.5\sigma$; and (c) $\Delta = 2\sigma$, $\xi_D = 1.5\sigma$. The differences of grand potential ($\Delta\Omega = \Omega_{CA} - \Omega_{RD}$, red solid circles), potential energy ($\Delta E = E_{CA} - E_{RD}$, green open triangles) and entropy per polymer chain ($\Delta S_s = \Delta S_{s,CA} - \Delta S_{s,RD}$, blue open squares) between two "standard" RD and CA particle distributions are shown along two paths: (d) $\Delta = 0.5\sigma$ and varying $\xi_D$; (e) $\xi_D = 1.5\sigma$ and varying $\Delta$.

**C. Depletion and interfacial attraction in the polymer-large-nanoparticle mixture.**

We also investigated the large nanoparticles immersed in polymer matrix, the radius of which is set to be $R_c = 5\sigma$ ($\approx 1.22R_{g,0}$). Since the size of particles is much larger than the segment, we turn on the depletion effect, by controlling the parameter $\xi_D$. Larger $\xi_D$ implies more rigid polymer chain. Here, we focused on the competition between the depletion and the interfacial interaction in the structural variation. When $\varepsilon = -0.2$, the system is close to the boundary between RD and CA phases. Particle distributions for $\Delta = 0.5\sigma$ and $\xi_D = 0$, $\Delta = 0.5\sigma$ and $\xi_D = 1.5\sigma$, $\Delta = 2\sigma$ and $\xi_D = 1.5\sigma$ are shown in Fig. 4, (a)-(c), respectively. The particle distribution changes from RD to CA along the path of increasing $\xi_D$ at constant $\Delta = 0.5\sigma$. While, on the contrary, the distribution changes from CA to RD along the path of increasing $\Delta$ at constant $\xi_D = 1.5\sigma$. Hence, we have the conclusion that narrow depletion layer (or flexible polymer chains) and/or "long"-range weak polymer-particle attraction can facilitate the dispersion of large nanoparticles into polymer matrix.

We took the particle distributions of Fig. 4(a) and (b) as two "standard" RD and CA distributions and calculate the differences of $\Delta\Omega = \Omega_{CA} - \Omega_{RD}$ , $\Delta E = E_{CA} - E_{RD}$ and $\Delta S_s = \Delta S_{s,CA} - \Delta S_{s,RD}$ along the two paths of varying $\xi_D$ with constant $\Delta = 0.5\sigma$ and varying $\Delta$ with constant $\xi_D = 1.5\sigma$ (Fig. 4(d) and (e)). Both $\Delta E$ and $\Delta S_s$ at all cases are positive, i.e., the aggregation of particles is energetically unfavored but entropically favored (depletion effect). In Fig. 4(d), the transition from RD to CA happens between $\xi_D = \sigma$ and $\xi_D = 1.5\sigma$ where $\Delta\Omega$ becomes zero. $\Delta S_s$ ($\Delta E$) increases (decreases) with $\xi_D$, implying that the increase of depletion layer (rigidity of polymer chain) enhances the aggregation of particles not only through the well-known depletion entropic effect but also through the alleviation of the penalty of interfacial energy. The transition from CA to RD in Fig. 4(e) happens at $\Delta$ slightly larger than $0.5\sigma$. The decrease of $\Delta S_s$ and the increase of $\Delta E$ with $\Delta$ imply that the contribution of enthalpy (entropy) in the structural variation is enhanced (weakened) with the increase of $\Delta$.

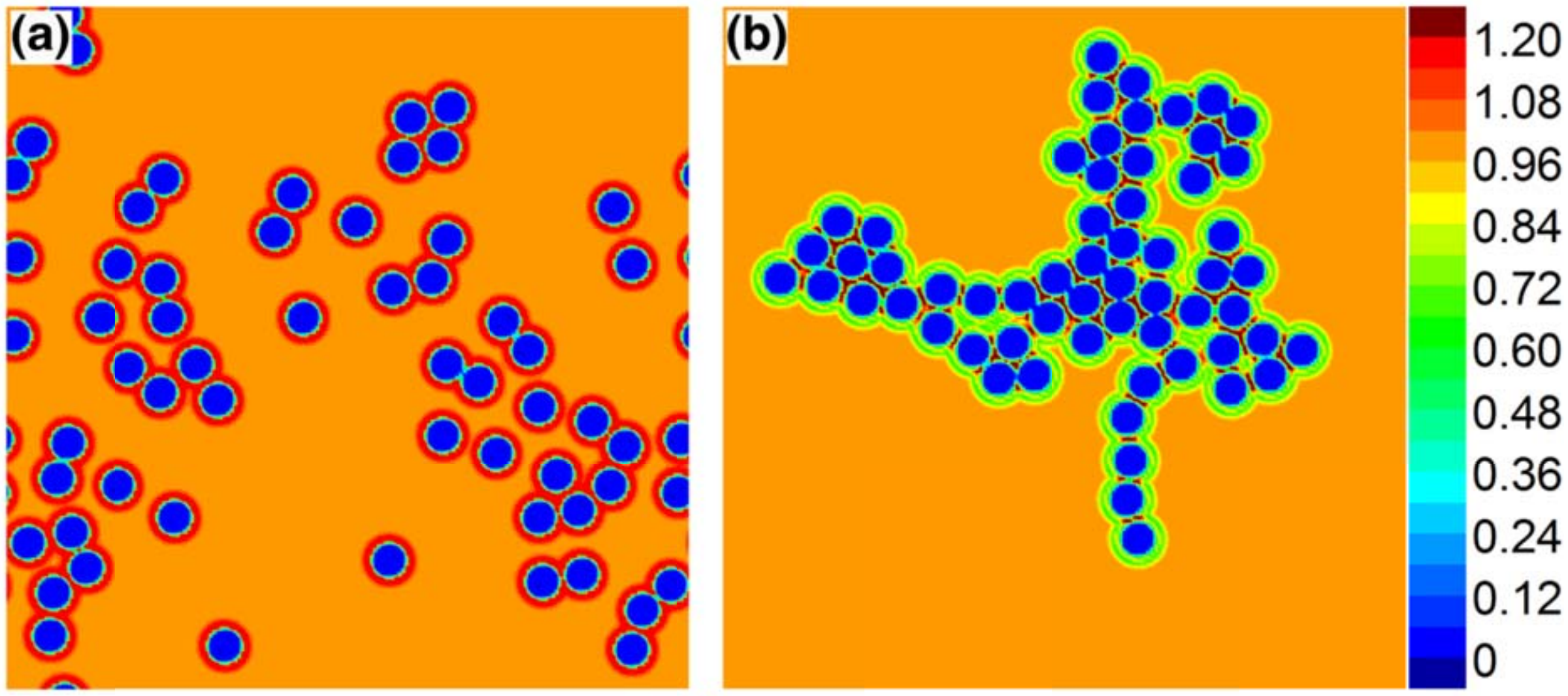

**FIG 5.** Structural variation of large nanoparticle/polymer composites in the case of strong attraction ($\varepsilon = -3$). (The system parameters are the same as in Fig. 4.) Two concentration plots of polymers are for (a) $\Delta = 0.5\sigma$, $\xi_D = 0$ and (b) $\Delta = 0.5\sigma$, $\xi_D = 2\sigma$.

When the attraction is strong ($\varepsilon = -3$), particles aggregate by bridging effect. In most cases (various $\xi_D$ and $\Delta$), the aggregation of large particles is loose, analogous to the LBA structure of small nanoparticles in Fig. 2(g). The typical density profile is shown in Fig. 5(a). Interestingly, we find the depletion effect can promote the bridging (**not contact**) aggregation of particles (Fig. 5(b)), when the depletion layer is much larger than the range of attraction ($\xi_D = 2\sigma >> \Delta = 0.5\sigma$). We set the particle distribution of Fig. 5(a) as the reference and get $\beta\Delta\Omega = -293 \pm 49$, $\beta\Delta E = -323 \pm 49$ and $\Delta S_s / k_B = -0.0634 \pm 0.0051$ for the formation of the structure in Fig. 5(b). These results imply that the closely bridging aggregation of large particles in Fig. 5(b) is enthalpically driven but entropically unfavorable, although it is triggered by the existence of a wide depletion layer around particles.

## IV. Conclusion

The field model of PNCs introduced in this work realizes the discrete description of particles which can predict the (more realistic) collective assembly of particles in the polymer matrix with careful consideration of the EV interactions, depletion effect and interfacial interaction. This model allows the investigation into the bridging and depletion effects on the multi-particle collective level instead of calculations based on two-particle correlations.[25, 26] It can reveal the entropic and enthalpic contributions in the variation of morphologies. Overall, it is a valuable approach to exploring and analyzing the rich mesostructures in polymer-based nanocomposites and is readily extendable to biology-related systems, such as dendrimer/drug encapsulation[46] and membrane/particle assembly.[47]

## Acknowledgements

This work was supported by the National Basic Research Program of China (973 Program) No. 2012CB821500 and the National Natural Science Foundation of China (NSFC) Nos. 21374073, 11074180, 21574096 and 21474074.

**APPENDIX: SCFT equations and implementation**

In our modle, the system of simple nanoparticle/polymer mixture is specified by nine parameters: $\sigma$ (segment size), $R_{g,0}$ (unperturbed radius of gyration of polymer chain), $R_c$ (radius of particles), $\xi_D$ (width of depletion layer), $\Delta$ (spatial range of polymer-particle interfacial interaction), $n_c$ (number of particles), $\mu$ (chemical potential of polymers), $\kappa^{-1}$ (dimensionless parameter proportional to the compressibility of the polymer matrix) and $\varepsilon$ (strength of polymer-particle interfacial interaction). We set $\sigma$ as the unit length and fix $R_{g,0}=4.08$ (N=100). $\mu$ is chosen that the volume/area fraction of bulk polymers is 1. $\kappa^{-1}$ varies with, such as, the concentration of the polymer solution or the temperature. Its value influences the results at a quantitative level. We set $\kappa^{-1}=3.33$ which corresponds to the incompressibility of polymethylmethacrylate (PMMA) melt at 450K.

SCFT is a very powerful approach in predicting the mesoscopic structures of multicomponent polymeric systems. In our model, we employ the grand-canonical description for the polymer solution and canonical description for nanoparticles. Given the interaction potentials [Eqs.(1), (2) and (5-7)], we follow the SCFT approach and obtain the mean-field grand potential:

$$\begin{aligned}\beta\Omega = \beta U_p + \beta U_c + \beta U_s + \beta U_A - \sigma^{-d}\int d\vec{r} w_p(\vec{r})\phi_p(\vec{r}) - \int d\vec{r} w_c(\vec{r})\rho_c(\vec{r}) \\ - z_p Q_p\left[w_p(\vec{r})\right] - n_c \ln\left\{Q_c\left[w_c(\vec{r})\right]/n_c\right\}\end{aligned} \tag{A1}$$

$w_p(\vec{r})$ and $w_c(\vec{r})$ are auxiliary fields to decouple interactions; $z_p = e^{\beta\mu}$ is the activity of polymers; $Q_p$ and $Q_c$ are the partition functions of a single polymer chain or nanoparticle in the auxiliary fields, respectively. The fields are given by

$$w_p(\vec{r}) = \kappa^{-1}\left[\phi_p(\vec{r}) + \phi_c^{eff}(\vec{r}) - 1\right]H_0(\vec{r}) + \kappa_h^{-1}\left[\phi_p(\vec{r}) + \phi_c(\vec{r}) - 1\right]H_h(\vec{r}) + \int_{R_c \le |\vec{r}-\vec{r}'|} d\vec{r}'\varepsilon \exp\left[-\left(|\vec{r}-\vec{r}'| - R_c\right)/\Delta\right]\rho_c(\vec{r}') \tag{A2}$$

$$\begin{aligned} w_c(\vec{r}) = &\sum_{\vec{r}'} \sigma^{-d} l^d \kappa^{-1}\left[\phi_p(\vec{r}') + \phi_c^{eff}(\vec{r}') - 1\right] Th(\xi_D) H\left(|\vec{r}-\vec{r}'| - R_c\right)\delta\left[\phi_{c,\vec{r}}(\vec{r}'), \phi_c^{eff}(\vec{r}')\right]H_0(\vec{r}') \\ &+ \sigma^{-d}\int d\vec{r}'\kappa^{-1}\left[\phi_p(\vec{r}') + \phi_c^{eff}(\vec{r}') - 1\right]Th(0)\delta\left[\phi_c(\vec{r}'), \phi_c^{eff}(\vec{r}')\right]H_0(\vec{r}') \\ &+ \sigma^{-d}\int d\vec{r}'\kappa_h^{-1}\left\{\left[\phi_p(\vec{r}') + \phi_c(\vec{r}') - 1\right]Th(0)H_h(\vec{r}') + V_{op}\left(|\vec{r}-\vec{r}'|\right)\rho_c(\vec{r}')\right\} \\ &+ \sigma^{-d}\int_{R_c \le |\vec{r}'-\vec{r}|} d\vec{r}'\ \varepsilon \exp\left[-\left(|\vec{r}'-\vec{r}| - R_c\right)/\Delta\right]\phi_p(\vec{r}') \\ &+ \sigma^{-d}\lambda\left\{\left[\rho_c(\vec{r}) - \rho_l\right]H\left[\bar{\rho}_c - \rho_c(\vec{r})\right] + \left[\rho_c(\vec{r}) - \rho_u\right]H\left[\rho_c(\vec{r}) - \bar{\rho}_c\right]\right\} \end{aligned}$$

(A3)

where

$$H_0(\vec{r}) \equiv H\left[\phi_{hc} - \phi_c(\vec{r})\right] + H\left[\phi_c(\vec{r}) - \phi_{hc}\right]H\left[1 - \phi_p(\vec{r}) - \phi_c(\vec{r})\right] \tag{A4}$$

$$H_h(\vec{r}) \equiv H\left[\phi_p(\vec{r}) + \phi_c(\vec{r}) - 1\right]H\left[\phi_c(\vec{r}) - \phi_{hc}\right] \tag{A5}$$

$$Th(x) \equiv 0.5\left\{1 - \tanh\left[3\left(|\vec{r}-\vec{r}'| - R_c - x\right)/\left(x+\sigma\right)\right]\right\} \tag{A6}$$

The function $\delta[...]$ in Eq. (A3) is the Kronecker delta function. The distributions of polymers and particles are given by

$$\phi_p(\vec{r}) = \frac{z_p N \sigma^d}{V}\int_0^1 ds q(\vec{r}, s) q(\vec{r}, 1-s) \tag{A7}$$

$$\rho_c(\vec{r}) = \frac{n_c}{VQ_c}\exp\left[-w_c(\vec{r})\right] \tag{A8}$$

The chain propagator $q(\vec{r}, s)$ satisfies the diffusion equation:

$$\frac{\partial q(\vec{r}, s)}{\partial s} = R_{g,0}^2 \nabla^2 q(\vec{r}, s) - N w_p(\vec{r}) q(\vec{r}, s) \tag{A9}$$

with the initial condition $q(\vec{r}, 0) = 1$. The partition functions of a single polymer chain or nanoparticle are given by

$$Q_p\left[w_p\left(\bar{r}\right)\right]=\frac{1}{V}\int d\bar{r}q\left(\bar{r},1\right) \tag{A10}$$

$$Q_c\left[w_c\left(\bar{r}\right)\right]=\frac{1}{V}\int d\bar{r}\exp\left[-w_c\left(\bar{r}\right)\right] \tag{A11}$$

The Eqs. (A1)-(A11) consists the set of SCFT equations. To find the numerical solutions, we first make a random initialization for the auxiliary fields and then update the densities or concentrations and auxiliary fields iteratively.[45] Pseudo spectral method is adopted to solve the diffusion equation.[44] *This iterative procedure is coupled with an "annealing" process (varying $\lambda$ in Eq.(7)) for the artificial potential, which controls the gradual formation of discrete particles.* The calculations were performed in two dimensions.